# 3D printed microfiber waveguide in C-shaped fiber for temperature and air pressure measurement

Qipeng Huang,[1] Shanmei Zeng,[1] Jingxian Cui,[2] Lin Htein,[2] Hwa-Yaw Tam,[2] Xuehao Hu,[3, *] and Hang Qu[1, *]

1) Department of Physics, Shantou University, 243 Daxue Rd. Shantou, China, 515063
2) Department of Electrical and Electronic Engineering, The Hong Kong Polytechnic University, Kowloon, Hong Kong SAR 997700, China
3) Department of Electromagnetism and Telecommunication, University of Mons

(*Electronic mail: xuehao.hu@umons.ac.be, haqux@stu.edu.cn)

In this study, we propose a microfiber waveguide for temperature and air pressure measurement. To improve mechanical strength of the sensor, a C-shaped fiber is sandwiches between two single mode fibers (SMFs) by fusion splice. The microfiber waveguide is 3D printed between two SMFs to connect two fiber cores by two-photon polymerization technology. Due to multimode guidance of this printed microfiber, a Mach-Zehnder interferometer (MZI) is obtained. This sensor exhibits a high temperature sensitivity of 361 pm/°C at 25°C to 45°C and a high air pressure sensitivity of 55 pm/kPa from 300hpa to 1000hpa. The MZI sensor features significant advantages such as small size, high stability, and easy fabrication, without the need for complex post-processing, showing great potential and broad application prospects in many sensing applications.

## I. INTRODUCTION

In recent years, gas pressure sensors have played a crucial role in a wide range of fields, including weather forecasting, industrial monitoring, biomedical instrumentation, and aerospace[1-3]. With the significant growth of the market demand for high-sensitivity pressure sensors, fiber-optic pressure sensors are gradually favored by more and more researchers because of their excellent performance in harsh environments[4,5]. Compared with traditional sensors, optical fiber sensors[6-10] are usually resistant to electromagnetic interference, corrosion resistance, high chemical stability, lighter size, and higher sensitivity because optical signals are transmitted through optical fibers and are affected by the environment of the sensing area composed of different optical fiber structures. Therefore, in the field of gas pressure measurement, the effective refractive index of the gas varies with the pressure, which makes fiber optic sensors a significant advantage in this field[11,12]. Given these advantages, fiber optic sensors are gradually replacing traditional sensors and are expected to become the dominant technology in this field in the future.

The structure of fiber optic pressure sensors is mainly composed of fiber Bragg gratings (FBGs)[4,5,13], Fabry-Perot interferometers (FPIs)[14-16], Mach-Zehnder interferometers (MZIs)[9,17], Sagnac interferometers[18], and sensors that use two different interference principles[19,20]. Among the many sensors, the FBG mainly uses the selectivity of wavelength to make the sensing device, although the structure is relatively simple, but its sensitivity is relatively low[21]. In contrast, an FPI operates based on the formation of interference in a resonator by incident light, which is highly accurate but the structure is usually more complex, so it is more difficult to make[14]. For sensors with multi-interference principles, for example, Yang et al. proposed a novel and compact fiber-optic gas pressure sensor using a chitosan-coated twin-core side hole fiber, with a pressure sensitivity of 589.41 nm/MPa and a temperature sensitivity of −0.47 nm/°C[19]. MZIs enable fiber optic sensing primarily by measuring the phase shift between two beams of coherent light, specifically the difference in the optical path length of the sensing arm and the reference arm can cause interference. Typically, the sensing arm is exposed to the external environment, while the reference arm is regulated unchanged. MZI-based sensors have shown significant advantages over other types of sensors in multi-parameter sensing applications due to their flexible structure, ultra-high parameter sensitivity, and excellent environmental robustness[22-26]. As a result, MZI fiber optic sensors offer unique technical features and advantages when it comes to pressure measurement. In recent years, many research results have emerged in the field of barometric pressure sensing, such as Hou et al. proposed a polydimethylsiloxane (PDMS)-sealed microfiber MZI with a compact length of 2.6 mm and a pressure sensitivity of 0.013 nm/kPa[27]. In 2023, Zhang et al. proposed an MZI based on tapered single-mode fiber and cascaded multimode fiber, which showed excellent pressure sensitivity of 5.751 nm/kPa[28]. However, in pursuit of high sensitivity, fiber optic MZI pressure sensors do have difficulty controlling their length. Therefore, how to further reduce the size and maintain high sensitivity has become a key issue in the development of fiber optic MZI pressure sensors.

Today, femtosecond lasers, with their ultra-short pulses and extremely high peak powers, have opened up a revolutionary avenue for the fabrication of advanced optical devices[29], especially in the fabrication of precision waveguide structures. At present, there are many methods for preparing femtosecond laser waveguides, such as distributed fiber gratings, and changing the refractive index and two-photon polymerization (TPP) printing waveguide technology in coreless fibers. Yin et al.[30] designed a three-dimensional four-channel filter using distributed fiber gratings, which achieved compactly with 50 nm spectrum spacing within 1450–1600 nm wavelengths. In coreless fibers, femtosecond lasers can write waveguide structures directly inside homogeneous coreless fibers[31]. By precisely controlling the laser focus to scan inside the fiber, small, permanent areas of refractive index change induced by the laser pulse create a channel for light transmission. This approach breaks the limitations of traditional optical fibers and provides unprecedented design freedom and integration potential for fiber optic photonics devices. On the other hand, femtosecond laser fabrication waveguides using two-photon polymerization technology have also developed rapidly in recent years, and have been widely used in life sciences, materials science, mechanics and other fields. TPP is a 3D printing technology with high manufacturing precision and flexibility, which enables the efficient fabrication of micro/nanostructures on fiber optic platforms, with excellent results in realizing complex high-performance optical, mechanical, and biological structures[32,33]. For example, G.E. et al[34] were the first to create a waveguide manifold on top of a four-core optical fiber tip to couple light into and from a single-core optical fiber, achieving this in a fast and low-cost manner. The results show that the insertion loss can be less than 5 dB. In the past decade of research, TPP has been widely used in the field of optics and photonics because of its characteristic size from nanometers to millimeters, and because of its ability to easily produce complex structures in 3D space, the manufactured sensors have small size in the micron range, high stability, and strong anti-interference ability.

In summary, integrating waveguide and TPP printing technology with temperature and pressure sensing could yield better results. Therefore, in this paper, we propose to employ TPP 3D printing technique to fabricate a microfiber waveguide in a C-shaped fiber that is fusion-spliced between two SMFs. This structure inherently constitutes a fiber MZI due to the multimode guidance of the printed microfiber. The diameter of the microfiber could be as small as to 1 μm. The C-shaped fiber constitutes mechanical support of the MZI as well as the sensing reservoir for gaseous or liquid samples. Using this route, one potentially could fabricate a MZI with any designated lengths and microfiber diameters, thus yielding better performance for spectral modulation and higher sensitivities for sensing applications. making the MZI structure highly robust, compact, and responsive.

## II. SENSING PRINCIPLE

The working principle of the sensor can be explained as follows: As shown in Figure 1, both sides of the optical fibers are standard single-mode fibers. When light enters the C-shaped fiber region through the single-mode fiber (SMF) at the input end, due to mode field mismatch, the fundamental mode (FM) is primarily coupled into the photoresist waveguide, while higher-order modes (HOM) are coupled into the air cavity. The light propagates through the photoresist in the waveguide structure and is then transmitted to the SMF at the output end. During this process, the HOMs traveling through the air are also coupled into the SMF at the output end, where they interfere with the FM propagating through the photoresist waveguide. This interference pattern can be understood as the result of the interaction between the FM and HOMs. In an MZI, the intensity of light is typically presented by the following formula[35]:

$$I = I_1 + I_2 + 2\sqrt{I_1 I_2} \cos \Phi \quad (1)$$

Among them, $I_1$ and $I_2$ represent the intensity of two-speed propagation light respectively, while the intensity I of the optical output of the optical fiber depends on the phase difference between the modes involved in the interference, which is expressed as:

$$\Phi = \frac{2\pi}{\lambda}(\delta n_{eff})L \quad (2)$$

In Eq. (2), the sensing length is represented by L, λ represents the wavelength, and $n_{eff}$ represents the refractive index difference between the FM and HOMs. Therefore, we can easily get that the expression of the final output light intensity of the SMF in this structure is:

$$I = I_{HOM} + I_{FM} + 2\sqrt{I_{HOM} I_{FM}} \cos[2\pi L(n_{eff}^{FM} - n_{eff}^{HOM})/\lambda] \quad (3)$$

where $I_{FM}$ represents the optical intensity propagating in the waveguide structure, and $I_{HOM}$ denotes the intensity of light propagating in the air cavity. Based on the theoretical formula, the refractive index of the medium during the propagation of the FM and HOMs can be expressed as: According to the theory of the above formula, the resonant wavelength ($\lambda_{peak}$) and the free spectral range (FSR) can be expressed as:

$$\lambda_{peak} = \frac{2\Delta n_{eff} L}{2m} \quad (4)$$

$$FSR = \frac{\lambda^2}{\Delta n_{eff} L} \quad (5)$$

In addition, the entire C-type fiber may form a Fabry-Perot cavity and produce interference. However, because the FSR of this interference is very small, it has almost no effect on the major interference pattern in the spectral range, so it can be ignored.

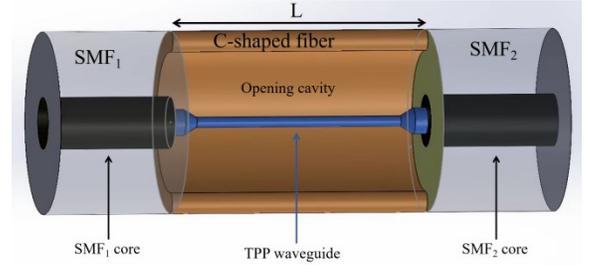

FIG. 1. Schematic diagram of the 3D structure.

## III. EXPERIMENT

In this study, we used the SMF-28e single-mode fiber and C-type fiber. The core diameter of the single-mode fiber is 8.2 μm, with a cladding diameter of 125 μm. The outer diameter of the C-type fiber is also 125 μm, while its inner diameter is 70 μm. We first selected a standard single-mode fiber jumper and used a fusion splicer (Fujikura, FSM-100P) to splice it with a C-type fiber. Next, the C-type fiber was cleaved to around 190-μm using a cleaver (Fujikura, CT50). Then, we spliced the other end of the C-type fiber with another standard single-mode fiber jumper. After splicing, the sample exhibited sufficient mechanical strength. Subsequently, the sample was placed onto the 6-axis motorized stage in the Femtosecond Laser uFAB Microfabrication Workstation (Newport), and photoresist (TPS-PA) was applied to immerse the whole C-shaped fiber. Following the pre-set patterning path, the microfiber could be printed inside the C-shaped fiber. The refractive index of the cured photoresist is 1.5385, compared to the air refractive index of approximately 1.0003[36]. Following development process in methyl ethyl acetate and a cleaning step with isopropanol, the completed MZI was obtained, as visualized in the micrographs in Figure 2. The waist region has a length of about 170 μm and a diameter of about 2 μm, with symmetrical tapered regions on either side. These 10-μm tapers increased in diameter to 8 μm, designed to facilitate a stable connection to the single-mode fiber.

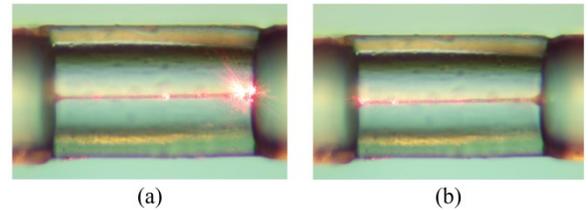

FIG. 2. Microscope images of the sensor while illuminated by a red laser pointer from the (a) right and (b) left side.

### A. Air pressure experiment

In our pneumatic experimental characterization, a vacuum pump was used to gradually reduce the pressure in the gas chamber from approximately 100 kPa (standard atmospheric pressure) to 30 kPa, with a step of 10 kPa, at a room temperature of 20 °C. As shown in Figure 3, we also used a single-mode fiber circulator and an optical fiber interrogator (HBK, FS22SI) to measure the spectrum. To characterize the gas refractive index measurement device, the sensor probe was placed inside a sealed gas chamber, where the internal pressure could be adjusted using an air compressor equipped with a built-in pressure gauge. Changing the pressure altered the refractive index of the air within the chamber.

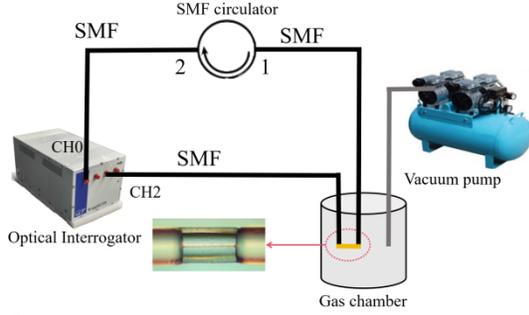

FIG. 3. Schematic diagram of the air pressure test setup.

As the pressure (and thus the refractive index) decreased, the spectrum exhibited a red shift. The results are shown in Figure 4, where (b) illustrates the linear fit curve of the interference peak shift versus pressure changes, along with error bars. The calculated sensitivity of the sensor is 55pm/kPa.

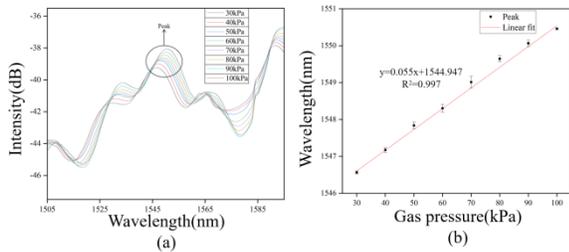

FIG. 4. For the barometric detection of the sensing structure, (a) is the peak drift, and (b) is the fitting curve with error bars.

We also measured the stability of the sensor under varying pressures, as shown in Figure 5. Measuring the spectrum for 300 seconds at 50 kPa and 90 kPa, with intervals of 60 seconds. It is evident that the sensor exhibits good air pressure stability.

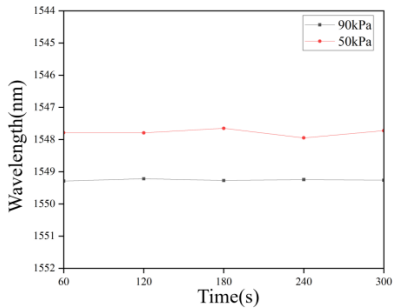

FIG. 5. Stability of the Air pressure test.

**B. Temperature experiment**

Figure 6 illustrates the experimental setup for the temperature experiment. We utilized a single-mode fiber circulator and an optical interrogator. The interrogator's output light is directed from port CH2 through the sensor's input section via an SMF, then connected to the single-mode fiber circulator (port 1 to port 2) at the sensor's output end, and finally returns to port CH0 of the Interrogator, allowing for direct analysis of the light input at port CH0. The sample is placed on a temperature-controlled breadboard (Thorlabs, PTC1) to maintain a constant temperature.

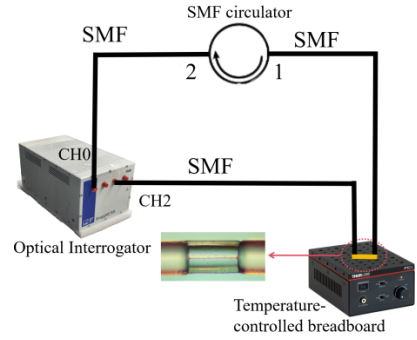

FIG. 6. Schematic diagram of the temperature test setup.

As the temperature rises, the interference spectrum exhibits a blue shift, as shown in Figure 7. In particular, (b) illustrates the linear fit curve of the interference peak offset versus temperature changes, with error bars. Based on this, the sensor's sensitivity is calculated to be -361pm/°C.

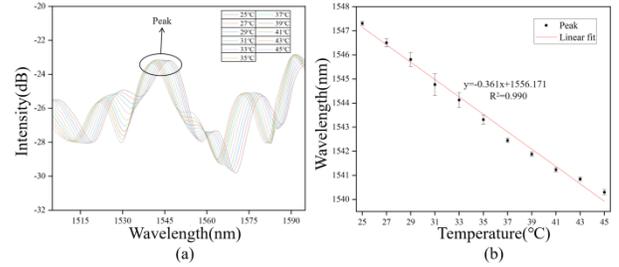

FIG. 7. For the temperature detection of the sensing structure, (a) is the peak drift, and (b) is the fitting curve with error bars.

We also measured the real-time response to temperature changes, as shown in Figure 8a. The red curve in the figure represents the temperature changes of the temperature-controlled breadboard. The third second marks the start of manual temperature control from 25°C to 27°C. The temperature-controlled breadboard undergoes a proportional-integral-differential (PID) process and stabilizes after about 60 seconds. We recorded a spectrum every second, and based on this, we determined that the temperature response time of this structure is approximately 1-2 seconds.

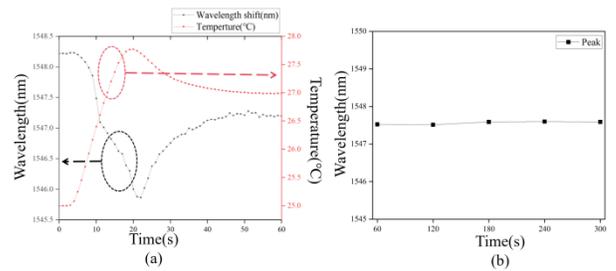

FIG. 8. Response time and stability test. The red dot in (a) is the real-time temperature change of the temperature-controlled breadboard, the black dot is the drift corresponding to the peak, and (b) is the stability of the temperature test.

We also measured the stability of the sensor under ambient temperature changes. As shown in Figure 8b, the spectrum was recorded from 25 °C to 27 °C for 60 seconds after the start of stability, and the spectrum was recorded every 60 seconds until 300 seconds. It can be seen that the temperature stability of the sensor is good.

Finally, it is worth noting that sensitive materials can be coated on the tapered region of the waveguide to enhance sensitivity. When the tapered region acts as a sensing element, it can stimulate the cladding mode, thereby enhancing the evanescent wave. The evanescent wave leaks into the cladding and interacts with the sensing environment, generating the output transmission spectrum. To further improve the sensor's performance, different parameters such as the waist length, and shape of the tapered geometry play a crucial role. Sensors based on the tapered configuration offer advantages such as low loss, large evanescent field, high sensitivity, strong confinement, miniaturization, and fast response time, showing great potential in various sensing applications.

## IV. CONCLUSION

The sensor proposed in this paper benefits from the designed miniature tapered waveguide structure, which has high temperature and pressure sensitivity of 356 pm/°C and 56 pm/kPa. In summary, the interferometric sensor is the most important device for measuring the parameters of the surrounding environment. MZI and MI sensors are widely used in sensing applications due to their high measurement accuracy, sensitivity, and fast response times. Compared to other sensors, these sensors also offer advantages such as high coupling efficiency, easy alignment, and low cost. Additionally, the tapered fiber geometry can be utilized for biomedical sensing or devices. The combination of various taper geometries in optical fibers can achieve interferometric sensors with higher sensitivity to parameters such as refractive index and temperature. The MZI sensors discussed all exhibit unique advantages, including small size, high stability, and ease of manufacturing, without requiring any additional complex processes. We can confidently conclude that MZI-based tapered fiber sensors have significant potential in large-scale sensing applications and a bright future.